\documentclass[twocolumn,aps,superscriptaddress,nofootinbib]{revtex4-1}
\usepackage{latexsym,amssymb,amsmath,epsfig,bm,psfrag}
\usepackage{color}
\usepackage[bookmarksnumbered,bookmarksopen,colorlinks,citecolor=red,linkcolor=blue]{hyperref}
\usepackage{graphicx}
\usepackage{times}
\usepackage{ulem,slashed}

\hypersetup{
    colorlinks=true,
    linkcolor=blue,
    filecolor=magenta,
    urlcolor=cyan,
    pdftitle={Sharelatex Example},
    bookmarks=true,
}

\newcommand{\Fig}[1]{Fig.~\ref{#1}}
\newcommand{\Eq}[1]{Eq.~(\ref{#1})}

\newcommand\be{\begin{equation}}
\newcommand\ee{\end{equation}}
\newcommand\bra{\langle}
\newcommand\ket{\rangle}

\def\Tef{T_{\rm eff}}
\def\Vef{V_{\rm eff}}
\def\mpt{\langle p_T\rangle}

\newcommand{\trento}{{T$_{\rm R}$ENTo }}

\begin{document}

\title{Volume effect on the extraction of speed of sound in high-energy nucleus-nucleus collisions}

\author{Jing-An Sun}
\email{jasun22@m.fudan.edu.cn}
\affiliation{Institute of Modern Physics, Fudan University, Shanghai 200433, China}
\affiliation{Department of Physics, McGill University
3600 rue University Montreal, QC Canada H3A 2T8}

\author{Li Yan}
\email{cliyan@fudan.edu.cn}
\affiliation{Institute of Modern Physics, Fudan University, Shanghai 200433, China}
\affiliation{Key Laboratory of Nuclear Physics and Ion-beam Application (MOE), Fudan University, Shanghai 200433, China}

\begin{abstract}

The determination of the speed of sound in quark-gluon plasma is a crucial aspect of understanding the properties of strongly interacting matter created in relativistic heavy-ion collisions. In this study, we investigate the impact of initial-state fluctuations on the extraction of the speed of sound in a quark-gluon plasma in the ultra-central collisions. By employing the \trento model for simulating initial conditions, we demonstrate that these fluctuations lead to sizable volume effect, which in turn corrects the measured values of the speed of sound. With respect to realistic conditions of high-energy heavy-ion experiments, we provide quantitative estimate of these corrections.

\end{abstract}
\maketitle

\section{introduction}

The speed of sound, $c_s$, defined as the ratio of the variation of pressure to that of energy density, relies on certain thermodynamic conditions,
\be
c_s^2 \equiv \frac{d P}{de}\Big|_X\,,
\ee 
where $X$ denotes the conditions of, for instance, constant temperature, constant entropy, etc.
In particular, with respect to a quantum chromodynamic (QCD) system in thermal equilibrium, the solution of the speed of sound from the lattice calculations of QCD implicitly assumes adiabatic condition~\cite{HotQCD:2014kol}.

Speed of sound can be measured in experiments. Consider, for instance, ideally a 
thermal system in equilibrium is constrained in a box with fixed volume. As one warms up the system by generating 
entropy {\it adiabatically}, effectively the increase of temperature, $\Delta T$, and entropy $\Delta S$, can be used to estimate the speed of sound. By definition and the thermodynamic relations~\cite{VANHOVE1982138}, 
\be
\label{eq:cs2}
c_s^2 \equiv \frac{d P}{de} = \frac{sdT}{Tds} =  \frac{\Delta T/T}{ \Delta S/S}\,.
\ee
In obtaining the last equation, one has taken into account the property of spatial homogeneity owing to the global thermal equilibrium condition, and the fixed volume condition: $\Delta V/\Delta S=0$.

Recently, in spirit of \Eq{eq:cs2}, the measurement of the speed of sound of a thermal QCD system has been proposed~\cite{Gardim:2019xjs} and carried out 
via the high-energy ultra-central collisions at the Large Hadron Collider (LHC)~\cite{CMS:2024sgx,ALICE_new}. In these ultra-central collisions, system volume 
approaches maximum, while event-by-event quantum fluctuations from nucleon-nucleon collisions provide a means of entropy production adiabatically.
In \Eq{eq:cs2}, 
the condition of thermal equilibrium is essential, although in realistic high-energy heavy-ion collisions, thermal equilibrium in the produced quark-gluon plasmas (QGP) is expected only locally, and even approximately, leading to a dissipative fluid description of the system expansion~\cite{Shen:2020mgh}. Accordingly, the conservation of energy and entropy allows one to effectively convert the system into a homogenous QGP with volume $V_{\rm eff}$ and temperature $\Tef$. Regarding final-state observables, hydrodynamic response relations are introduced to further relate the effective temperature $\Tef$ to the mean value of the transverse momentum of the produced particles $\mpt$: $\Tef\propto \mpt$, and the total entropy to the charged particle multiplicity: $S\propto dN_{\rm ch}/dy$. Eventually, the conjecture~\cite{Gardim:2019xjs} states that the speed of sound in QGP can be measured in experiments from the ultra-central collisions by 
\begin{align}
\label{eq:cs2_hic}
(c_s^0)^2 = \frac{\Delta \ln \Tef}{\Delta \ln S} = \frac{\Delta \ln \mpt}{\Delta \ln(dN_{\rm ch}/dy)}\,.
\end{align}
To avoid confusion, in \Eq{eq:cs2_hic} and the following, we denote the extracted speed of sound from the ratio of relative change of the effective temperature to that of the total entropy by $c_s^0$.

To validate this conjecture, several theoretical analyses have been performed~\cite{Gardim:2019brr,Gardim:2024zvi,SoaresRocha:2024drzJF,Nijs:2023bzv}. 
In this work, assuming the well-established hydrodynamic response relations, we examine the extraction of the speed of sound with respect to the fluctuating initial state of realistic heavy-ion collisions. On the first sight, it might seem trivial because the first equation in \Eq{eq:cs2_hic} follows simply the thermodynamic relation in \Eq{eq:cs2}. However, on an event-by-event basis, initial state density with fluctuations is neither homogeneous, nor sufficiently large that variation of the effective volume can be negligible. As a consequence, finite-volume effect arises which corrects the extracted speed of sound from \Eq{eq:cs2_hic}, leading to,
\be
\label{eq:cs2V}
c_s^2 = \frac{dP}{de} = \left(1 - \frac{\Delta \ln \Vef}{\Delta \ln S}\right)^{-1} \frac{\Delta \ln \Tef}{\Delta \ln S}\,.
\ee
Here, the ratio between the relative change of volume and total entropy $\Delta\ln \Vef/\Delta \ln S$, characterizes the response of the system volume as entropy is produced, which is not necessarily vanishing as one converts a fluctuating system to a homogeneous one. 
Note that, 
the condition $\Delta\ln \Vef/\Delta \ln S=0$ implicitly states that there is no work being done to a system, as the system energy or entropy is varied during the measurement of the speed of sound, there must be heat transfer, hence adiabiticity is violated.  More detailed discussions can be found in Appendix.~\ref{app:A}.

\section{Effective modeling of initial state}

Let us first briefly review the \trento model that we shall use.
\trento Model~\cite{Moreland:2014oya} provides one of the state-of-the-art numerical modelings of the initial stage of relativistic nucleus-nucleus collisions, with several parameters that represent different physics aspects. There are three substantial ingredients in the model, 
\begin{itemize}
\item A Gaussian beam-integrated proton density function with width $w$,
\be
\int dz \rho(x,y) \sim \exp\left(-\frac{x^2+y^2}{2w^2}\right)\,.
\ee
In principle, the width captures the size of a proton as it participates the collisions.
\item A fluctuating thickness function for each proton, $T_{A,B}(x,y)\sim \alpha_{A,B} \int dz \rho$, with $A$ ($B$) labeling the pair of projectiles and $\alpha_{A,B}$ random weights satisfying a gamma distribution with unit mean, 
\be
P_k(\alpha) = \frac{k^k}{\Gamma(k)} \alpha^{k-1} e^{-k\alpha}\,.
\ee
Here, the fluctuating parameter $k$ is also inversely proportional to the variance. Owing to the appearance of the random weights, the resulted multiplicity receives extra fluctuations. Larger fluctuations are realized via smaller values of the fluctuation parameter $k$, and vice versa.  
\item A reduced thickness function that characterizes initial state entropy density,
\be
\label{eq:rTh}
s(\tau_0,x,y)= C T_R(x,y) = \left(\frac{T_A^p + T_B^p}{2}\right)^{1/p}\,,
\ee
where the generalized mean parameter $p$, is introduced to effectively average the thickness functions from the two colliding nucleons. The overall factor $C$ can be adjusted, according to the total multiplicity in collisions.
\end{itemize}

The three parameters, width $w$, fluctuation parameter $k$, and the generalized mean $p$, are allowed to vary to reproduce the initial entropy density profile of a thermalized QGP, on different grounds. For instance, $p=1$ gives rise to arithmetic average in \Eq{eq:rTh}, which leads to the Monte Carlo wounded nucleon ansatz~\cite{Miller:2007riGlauber} and deposits a blob of entropy for each nucleon. For $p=0$, entropy is deposited around the center of the nucleon-nucleon collision, and the resulted density profile approximates that obtained in the IP-Glasma model~\cite{Schenke:2012wbIP}. With respect to a number of observables in experiments, these parameters have been determined based on Bayesian analyses~\cite{JETSCAPE:2020mzn}, with the most probable values found in ranges, $p\in [0.063,0.139]$, $w\in[0.81,1.19]$ fm and $k\in[0.98,1.05]$. In the current study, we consider $w=1$ fm, $k=1$, and $p=0$ as the most probable values.

\section{Extraction of the speed of sound}

\begin{figure}[t]
\begin{center}
\includegraphics[width=.50\textwidth] {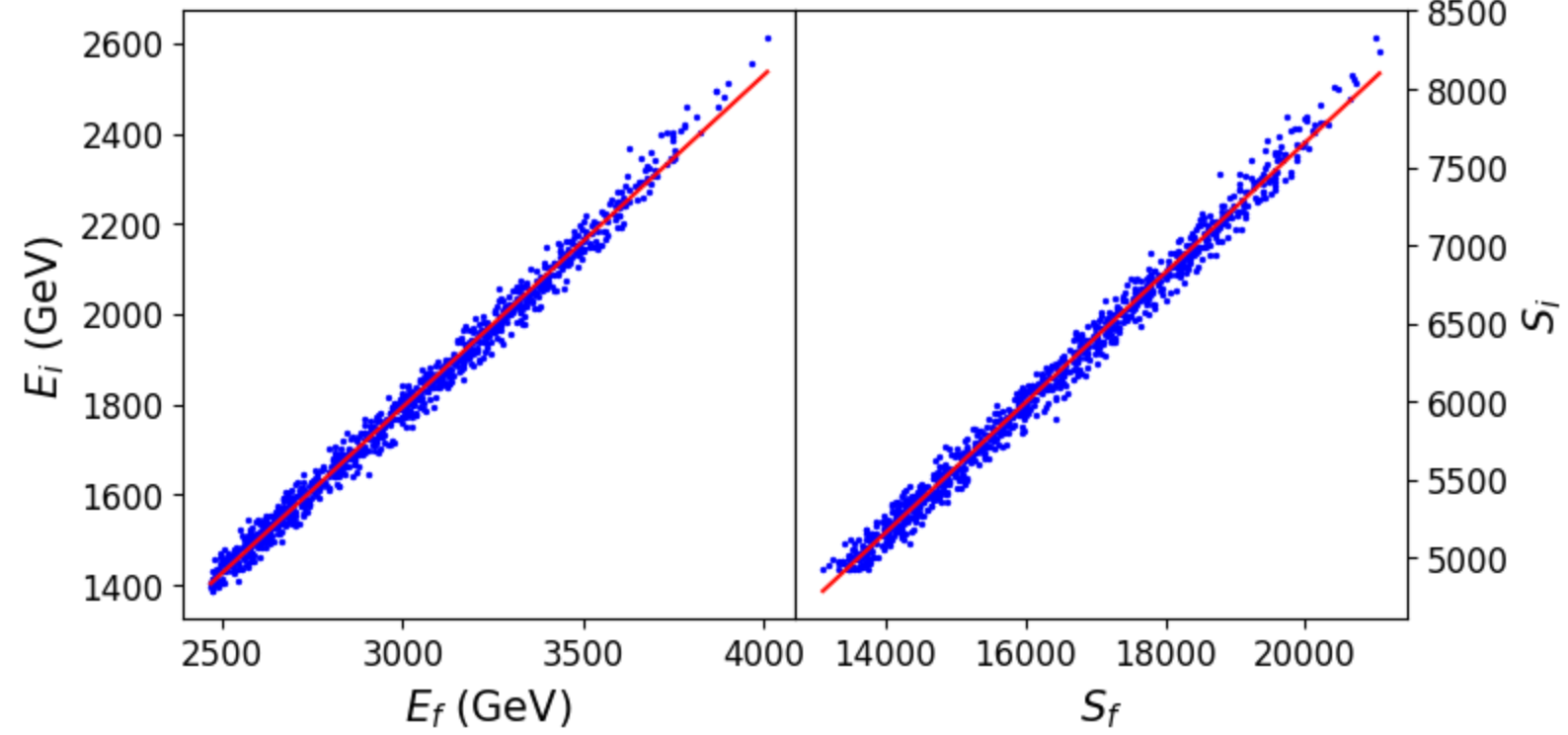}
\caption{
\label{fig:ic_final} The linear correlations between initial- and final-state energy (left panel) and entropy (right panel). 
}
\end{center}
\end{figure}

\begin{figure*}[t]
\begin{center}
\includegraphics[width=1.0\textwidth] {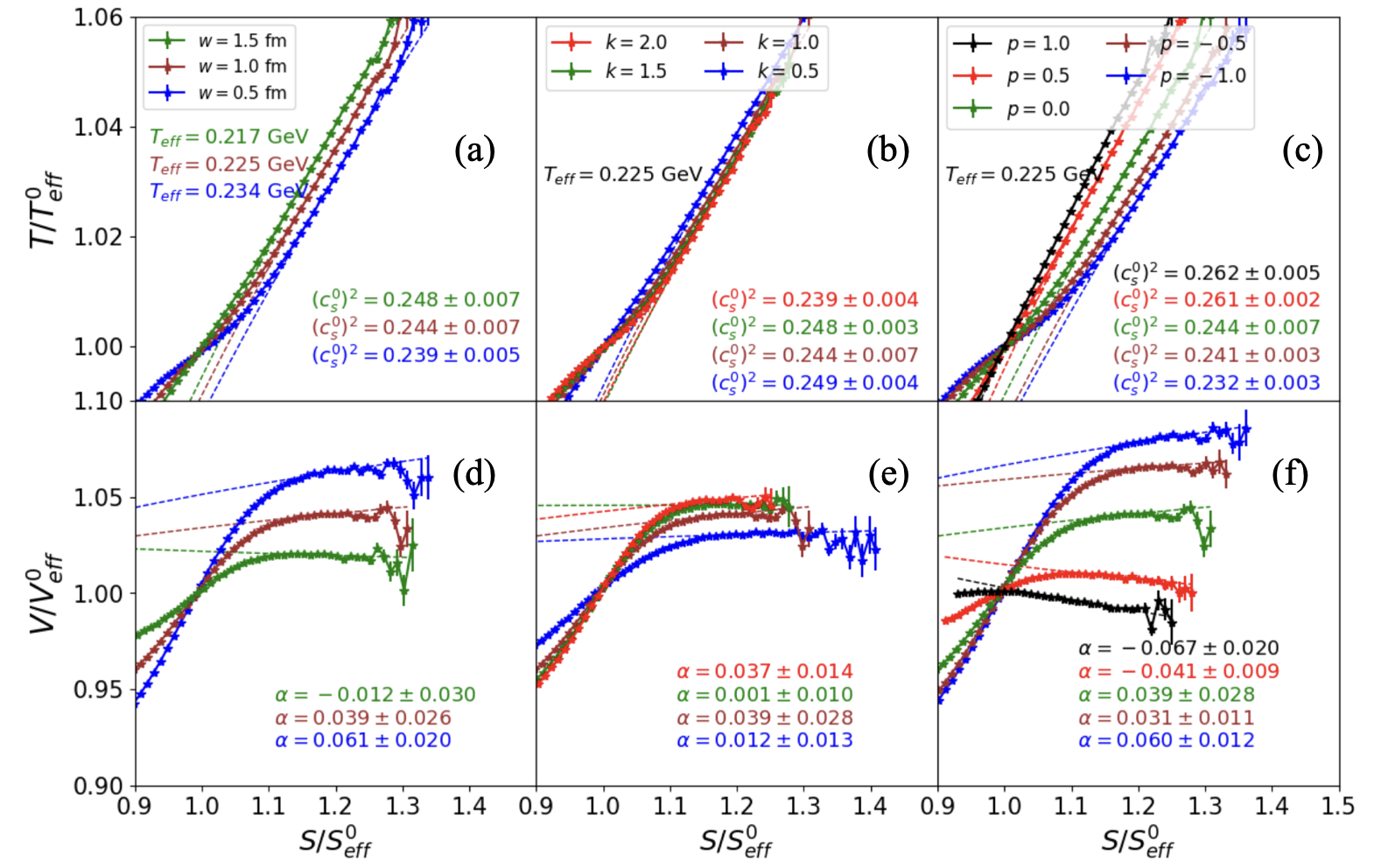}
\caption{
\label{fig:fig1} The initial-state parameter dependent relations between $\Tef$ and $S$, and $\Vef$ and $S$. Left panels ((a) and (b)): variation of width parameter $w$, with $p$ and $k$ fixed. Middle panels ((c) and (d)): variation of the fluctuation parameter $k$, with $p$ and $w$ fixed. Right panels ((e) and (f)): variation of the fluctuation parameter $p$, with $k$ and $w$ fixed. 
Note that the points associated with temperatures below the cross-over temperature actually capture systems of hadron gas.}

\end{center}
\end{figure*}
The procedure of extracting the speed of sound in the QGP medium starts with solving the effective temperature and effective volume~\cite{Gardim:2019xjs} from the equations,
\begin{subequations}
\label{eq:ESf}
\begin{align}
&E = E_{\rm final} = e(\Tef) V_{\rm eff}\,,\\
&S = S_{\rm final} = s(\Tef) V_{\rm eff}\,,
\end{align}
\end{subequations}
where the final-state total energy and entropy are estimated from particles emitted from the freeze-out hyper-surface. 
A parameterized lattice equation of state (LEOS) that interpolates between the hadron resonance gas at low temperatures to the lattice results of QCD at high temperatures, relates the total energy and entropy to an effective temperature and an effective volume. With same parameters as those used in \cite{HotQCD:2014kol}, the effective parameterization of the LEOS has been widely applied in many hydrodynamic calculations.


The equations can be applied as well if the total energy and entropy are given by the initial states,
\begin{subequations}
\label{eq:ESi}
\begin{align}
E = E_{\rm initial} &= \tau_0 \int d\vec x_\perp d\eta_s e(\tau_0,\vec x_\perp, \eta_s) \,,\\
S = S_{\rm initial} & = \tau_0 \int d\vec x_\perp d\eta_s s(\tau_0,\vec x_\perp, \eta_s)\,,
\end{align}
\end{subequations}
where the initial energy density is achieved by converting the entropy density using the same LEOS, $e=e(s)$\footnote{
An alternative procedure is to relate the reduced thickness $T_R(x,y)$ to the initial energy density profile, then entropy density is obtained by a conversion via LEOS $s=s(e)$~\cite{JETSCAPE:2020mzn}. 
}.
With respect to the total entropy, we are allowed to define centrality and to focus on the ultra-central events.
Although the resulted extraction of the speed of sound is not identical between initial state and the final state, the two speeds are linked with each other as there exist linear correlations between initial state and final state, as shown in \Fig{fig:ic_final}.
Accordingly, it is still valuable to solve the speed of sound for the initial state, for which the analysis applies to a static thermal fireball subject to the QCD equation of state, in the presence of event-by-event fluctuations.

The initial proper time $\tau_0$ in hydrodynamic modeling is often considered with respect to a quick thermalization assumption, with $\tau_0\leq 1$ fm/c. In our analysis, the exact value of $\tau_0$ is irrelevant to the extraction of the speed of sound. 
The dependence on the space-time rapidity, $\eta_s={\rm arctanh}(z/t)$, reflects the breaking of boost invariance in high-energy collisions, which is necessary for the conservation of energy during medium expansion. In this work, we consider a factorized density and a two-parameter 
function $s(\tau_0,\vec x_\perp, \eta_s)=s(\tau_0,\vec x_\perp)\times w(\eta_s)$ for the longitudinal density profile~\cite{Chatterjee:2017ahy}, 
\be
w(\eta_s) = \exp\left(-\theta(|\eta_s|-\eta_M)\frac{(|\eta_s|-\eta_M)^2}{2\sigma_\eta^2}\right)\,.
\ee
Varying the parameters, $\eta_M$ and $\sigma_\eta$, the shape of the longitudinal distribution can be adjusted accordingly. 
In particular, the shape affects the entropy and energy in the mid-rapidity region, $\eta_s\in [-0.5,0.5]$, and potentially plays a role in the rapidity selection in the extraction of the speed of sound in experiments~\cite{Nijs:2023bzv,ALICE_new}. We nevertheless leave the discussion on the $\eta_s$ dependence to future works. 

Ideally, with respect to the energy and entropy conservation, \Eq{eq:ESf} and \Eq{eq:ESi} are equivalent. However, it should be emphasized that, due to the finite dissipative effect in realistic QGP, from \Eq{eq:ESi} the resulted relation between the effective temperature and entropy is not expected comparable to the relation between the mean transverse momentum and charged particle multiplicity in experiements\footnote{Finite viscous effects lead to entropy production and a change in the mean transverse momentum.  Accordingly, in the ultra-central collisions, the distributions of the mean transverse momentum and the charged particle multiplicity, are supposed to be distorted, in comparion to the distribution of $\Tef$ and $S$ solved from \Eq{eq:ESi}.
}.

By imposing these identities in \Eq{eq:ESf}, as well in \Eq{eq:ESi}, the medium from realistic heavy-ion collisions is effectively converted to a homogeneous QGP in equilibrium, with volume $V_{\rm eff}$ and temperature $\Tef$.  In order to solve a speed of sound that is consistent with LEOS, the conversion must be adiabatic. Accordingly, volume between the two systems must differ.
For instance, for inhomogeneous systems extra entropy is produced during the conversion, similar to the thermalization process. If there is no change of volume between the realistic QGP and the homogeneous QGP, without work there must be heat flow out of the system according to the first law of thermodynamics, thus adiabaticity condition is violated. 


\subsection{Dependence on initial state model parameters}

\begin{figure*}[t]
\begin{center}
\includegraphics[width=1.0\textwidth] {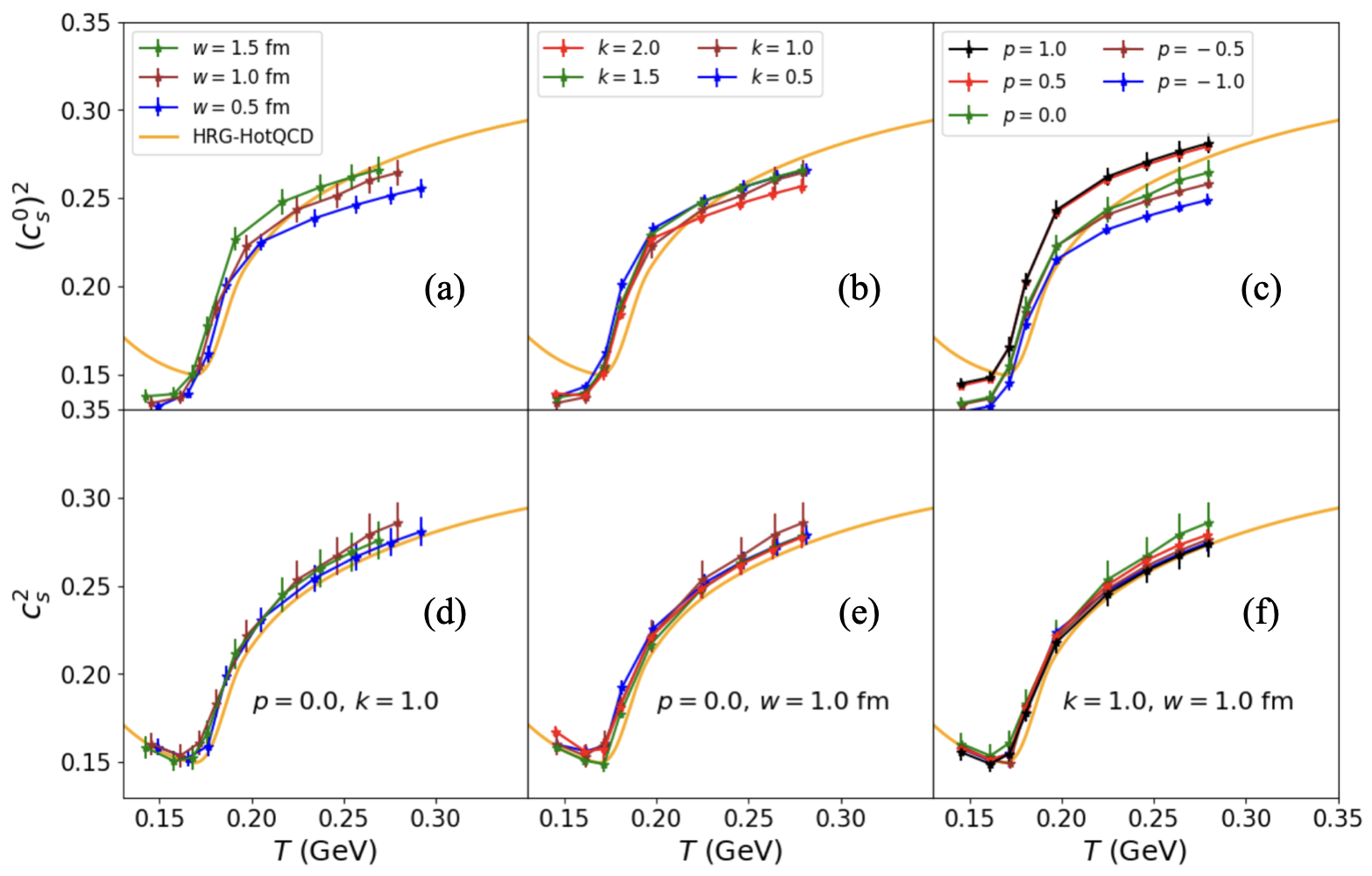}
\caption{
\label{fig:cs2} Extracted speed of sound from directly the linear relation between $\Tef$ and $S$, $(c_s^0)^2$ (top panels), and the extraction with corrections from the volume effect, $c_s^2$ (bottom panels), as a function of the effective temperature. The orange line corresponds to the sound speed from LEOS.
Dependence on the initial-state parameters, $w$, $k$, and $p$, are obvious in $(c_s^0)^2$ in (a), (b), and (c), but significantly suppressed when volume effect is corrected, as in $c_s^2$ in (d), (e) and (f). 
}
\end{center}
\end{figure*}

Without solving the entire medium evolution towards particlization, it is significantly simplified to extract the speed of sound from \Eq{eq:ESi} on an event-by-event basis. In particular, the effect of initial state fluctuations on the extraction of the speed of sound can be studied.
For each set of parameters in the \trento model, $(w,k,p)$, we select and simulate roughly five million events across the 0\% to 5\% centrality class of Pb-Pb collisions. We also vary the overall constant $C$ in \Eq{eq:rTh}, in order to scan a wider range of effective temperatures~\cite{Gardim:2024zvi}.  

We first fix the fluctuation parameter $k$ and the generalized mean $p$ by their most probable values, and vary the effective nucleon width around $w=1.0$ fm, results are shown in the left panels of \Fig{fig:fig1} ((a) and (d)). 
In analogy to the measurements of mean transverse momentum and multiplicity in experiments, we plot the effective temperature normalized by its average value in these ultra-central events, $\Tef/\Tef^0$, as a function of the total entropy scaled by its corresponding average, $S/S^0$. 
As shown in \Fig{fig:fig1} (a), these are data points solved within approximately 40 bins in the 0\%-5\% centrality class. Statistical errors of these points are estimated with respect to their standard deviation. At $S>S^0$, the system size tends to saturate with a vanishing impact parameter, one observes a perfect linear relation between $\Tef/\Tef^0$ and $S/S^0$, with the slope corresponding to $c_s^0$ 
that can be extracted following \Eq{eq:cs2_hic}. 
Note that, unlike the mean transverse momentum, from event-by-event simulations of initial states, there is no plateau and dip observed below $S^0$ in the effective temperature~\cite{ALICE_new,CMS:2024sgx,Nijs:2023bzv}. The plateau and dip structure which reflects the residual dependence on the system size of the mean transverse momentum, disappears as the system size saturates when centrality approach a limit, e.g., $S>S^0$. Similar effect can be observed in $\Tef$ that violation of the linearity exhibits in the smaller entropy region. Nonetheless, the extraction of the speed of sound does not rely on these structures.
 With the width growing from $w=0.5$ fm (blue points), to $w=1.0$ fm (brown points), to $w=1.5$ fm (green points), the resulted effective temperature decreases, but the slope increases.

In \Fig{fig:fig1} (d), the variation of the effective volume is plotted similarly, as a function of total entropy, for the ultra-central collisions events. We notice that although the original size of the collision system saturates, as the impact parameter approaches zero for $S>S^0$, the effective volume grows linearly. 
The slope of these lines, which we indicate as $\alpha=\Delta\ln \Vef/\Delta \ln S$, exhibits a systematic decrease as the width increases from $0.5$ fm to $1.5$ fm.

We next focus on the effect of the fluctuation parameter $k$. Analogously, we scan $k$ around its most probable values, while $w$ and $p$ are fixed. The corresponding results are shown in the middle panels of \Fig{fig:fig1} ((b) and (e)).  We notice that variation of $k$ does not change the effective temperature, neither the total entropy. Consequently, the linear relation between $\Tef$ and $S$ is barely affected by $k$. Unless for the small value, $k=0.5$, namely, large multiplicity fluctuations, the effect of $k$ on the effective volume is negible as well. 
With respect to the variation of $k$, the linearity between $\Vef$ and $S$ in the ultra-central events is weakly affected. 

The results in \Fig{fig:fig1} (c) and (f) are associated with the effect of $p$, for which we repeat the simulations with fixed values of both $k=1.0$ and $w=1.0$ fm, but vary $p$ from -1 to 1. The increase of $p$ does not impact the effective temperature, but does lead to 
a significant enhancement in the total entropy and the effective volume. 
Accordingly, the extracted slope $c_s^0$ increases, and $\alpha$ decreases, as $p$ increases. 

Interestingly,  if one takes into account the linear relations $\Tef\propto \bra p_T\ket$ and $S\propto dN_{\rm ch}/dy$, the parameter dependence of $\Tef$ and $S$ found in our simulations are consistent to the observations from the Bayesian analyses of full hydrodynamic calculations~\cite{JETSCAPE:2020mzn}. First, the width parameter $w$ has a negative correlation to the effective temperature (or $\bra p_T\ket$), and a positive correlation to the total entropy (or $dN_{\rm ch}/dy$). The fluctuation parameter $k$ is barely correlated to both the effective temperature (or $\bra p_T\ket$) and the total entropy (or $dN_{\rm ch}/dy$). The generalized mean $p$ is positively correlated to the total entropy (or $dN_{\rm ch}/dy$), while it has negligible effect on the effective temperature (or $\bra p_T\ket$).

\subsection{Volume effect}

\begin{figure}
\begin{center}
\includegraphics[width=0.45\textwidth] {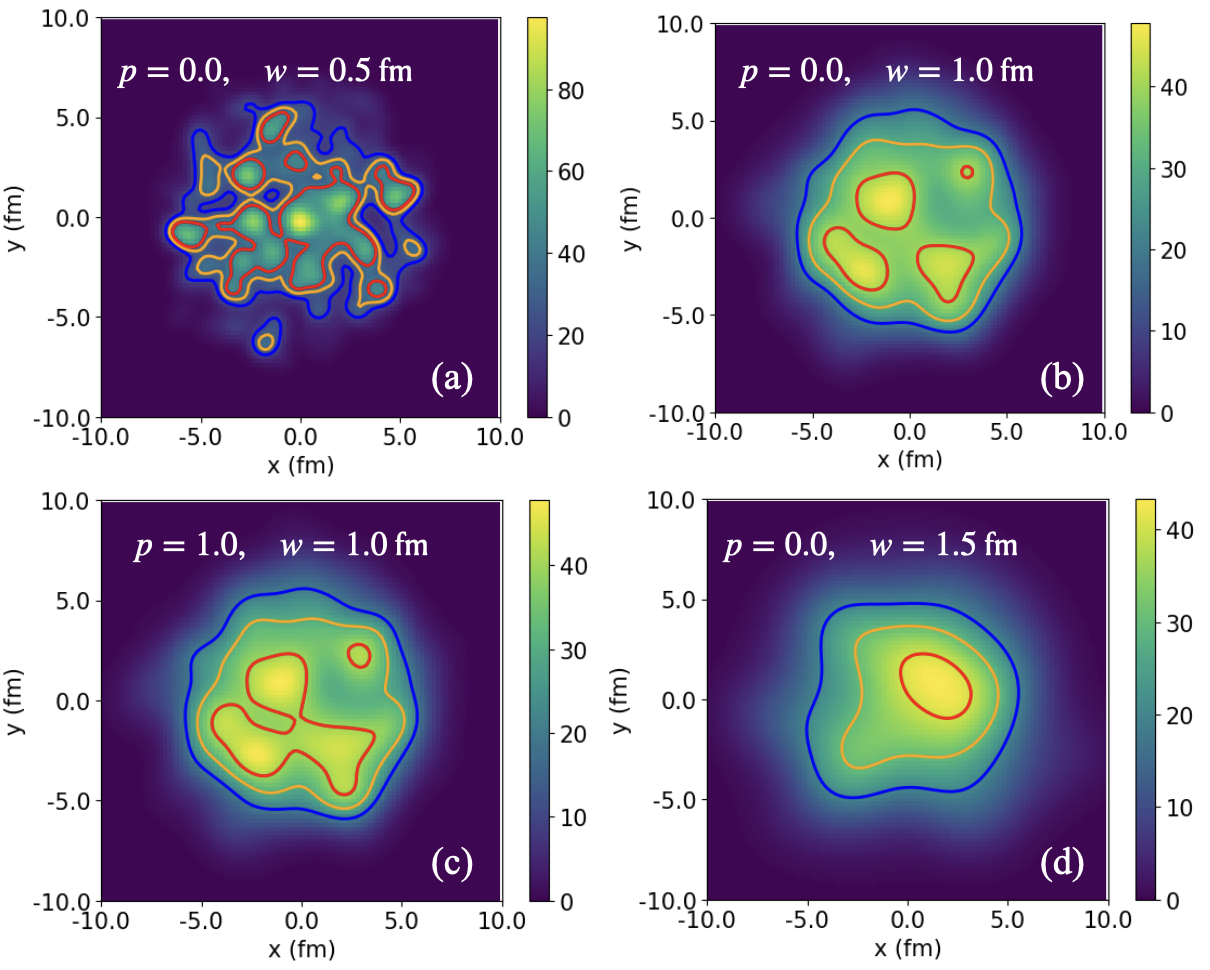}
\caption{
\label{fig:entropy} Initial-state entropy density generated from the \trento model with the same seed generator to demonstrate the change of inhomogeneity with respect to the increase of $p$ and/or $w$. 
}
\end{center}
\end{figure}

One immediately deduces that $c_s^0$ barely relies on the fluctuation parameter $k$, 
as both $\Tef$ and $S$ are not affected by the variation of $k$. 
However, the correlations of $\Tef$ and $S$ to $w$ and $p$, result in strong initial-state parameter dependence of the extracted speed of sound $c_s^0$ from the event-by-event simulations.
As shown in the top panels of \Fig{fig:cs2}, $c_s^0$, the slope extracted directed from the linearity between $\Tef$ and $S$, captures to a large extent the physical speed of sound, as comparing to the input values from LEoS. 
However, the exacted values of $c_s^2$ can deviate from the physical speed of sound, depending on $p$ and $w$. For instance, at $\Tef=225$ MeV, corresponding to the ultra-central PbPb collisions at $\sqrt{s_{NN}}=$ 5.02 TeV, small value of $p=-1$ gives rise to an underestimate of the speed of sound, while large $p$ leads to overestimation. In fact, one notices positive correlations between $(c_s^0)^2$, and $p$ and $w$. 

To extract the physical values of the speed of sound, $c_s$, a volume effect correction is required in addition to $c_s^0$.
To some extent, the volume effect arising from the event-by-event simulations is rooted in the spatial inhomogeneity of the initial state density profile. \Fig{fig:entropy} illustrates four typical initial-state density profiles generated by the \trento model. 
With both $p$ and $w$ increase (from (a) to (d)), 
instead of being discrete, the density distribution tends to be more smooth and connected. 
On the other hand, recall that the ratio $\alpha$ grows monotonically from being negative towards being positive, as $p$ and $w$ increase. In fact, it can be shown as in the Appendix, with respect to an inhomogeneous density, the change of the effective volume with respect to entropy production can be positive, if the density distribution is extremely discrete (as in panel (a) and (b) in \Fig{fig:entropy}), and can be negative if the density distribution is more connected (as in panel (c) and (d) in \Fig{fig:entropy}). Moreover, the absolute change of the effective volume is large when the entropy production leads to larger spatial inhomogeneity.

Because the effective volume $\Vef$ varies as the entropy increases, corrections due to the volume effect must be taken into account. According to \Eq{eq:cs2V}, the speed of sound extracted with respect to different parameters, $w$, $k$, and $p$ are shown in the lower panel of \Fig{fig:cs2}. With the volume effect correction, improvement is remarkable comparing to the LEoS results. 
It is worth noticing that, with the volume effect correction, the extraction also precisely captures the non-monotonic increase of the speed of sound at low temperatures, where the system becomes a hadron gas. 
Moreover, after the volume effect correction, parameter dependence of the extracted speed of sound becomes negligible. This is important, because the physical value of the extracted speed of sound must be independent of the initial-state parameters, as $c_s$ reflects only the thermodynamic nature of the medium.


\begin{figure*}[t]
    \centering
    \includegraphics[width=.90\linewidth]{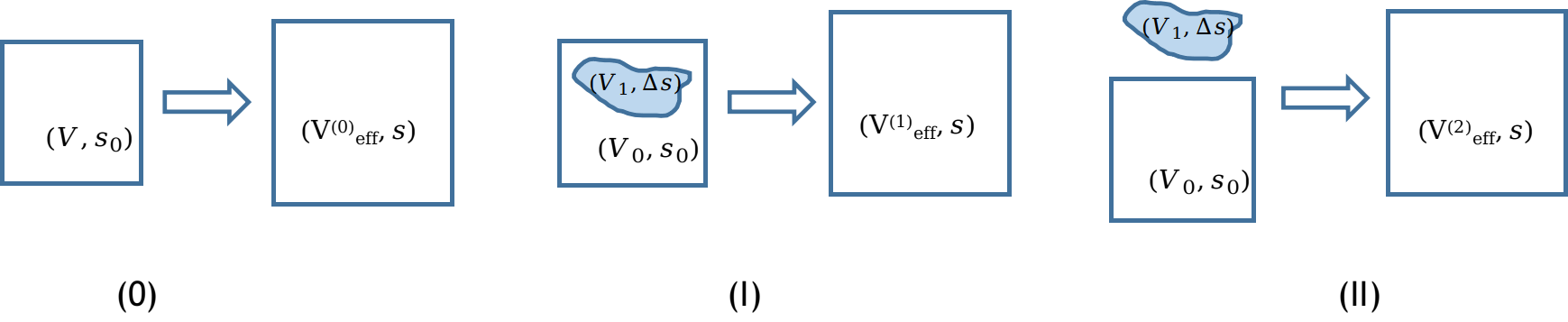}
    \caption{Inhomogeneous entropy production due to quantum fluctuations inside a system (shaded area) (I) or outside a system (shaded area) (II). When converting to a homogeneous thermal system with volume $\Vef$ and entropy density $s$, a finite ratio $\alpha = \frac{\Delta \ln V}{\Delta \ln S}$ is generated. }
    \label{fig:demo}
\end{figure*}

\section{Summary and discussion}

With the help of the \trento modeling of initial states, we are able to investigate the effect of event-by-event fluctuations on the extraction of the speed of sound proposed in Ref.~\cite{Gardim:2019xjs}. On one hand, these event-by-event fluctuations are crucial as they provide a means of variation of entropy in the systems created in ultra-central nucleus-nucleus collisions adiabatically. On the other hand, however, spatial inhomogeneity induced by these fluctuations results in variation of the effective volume. Accordingly, the extraction formula, which relates the speed of sound in QGP to the ratio of the relative change of mean transverse momentum and the relative change of the charged particle multiplicity, must receive corrections from the volume effect.

In fact, the variation of the effective volume is inevitable, as the conversion from an inhomogeneous density distribution to a fixed-volume QGP system can not be realized adiabaticity, it thus affects the extraction of the speed of sound nonetheless.
With the correction from the volume effect, the extracted speed of sound from our simulations appears physical, as they are not only quantitatively consistent with the LEOS results, but also independent of initial-state parameters.


\acknowledgements
We are grateful to Jean-Yves Ollitrault and Wei Li for helpful discussions. Jing-An Sun also thanks the hospitality of the nuclear theory group at McGill University, especially for the useful conversations with Charles Gale, Sangyong Jeon, Lipei Du, Xiangyu Wu, and Nicolas Miro Fortier. The computations are made on the Beluga supercomputer system at McGill University, managed by Calcul Québec and the Digital Research Alliance of Canada. This work is supported by the National Natural Science Foundation of China under Grant No. 12375133 and 12147101.

\appendix

\appendix

\section{Inhomogeneity and $\Delta \ln \Vef/\Delta \ln S$}
\label{app:A}

To illustrate the role of spatial inhomogeneity due to the quantum fluctuation of the entropy production in the ultra-central high-energy heavy-ion collisions in the volume effect, we consider a toy model, as shown in \Fig{fig:demo}.

Let us start with a trivial conversion of a homogeneous system with volume $V$ and entropy density $s_0=s(e_0)$ according to the equation of state, corresponding to the case (0). Conversoin with respect to the identification of total energy and entropy gives rise to an identical system, especially, $\Vef^{(0)} = V$. One may treat the case (0) as the background system without extra entropy production.

{\bf Case (I):} 
The quantum nature of the initial-state fluctuations in heavy-ion collisions implies inhomogeneous entropy production. For simplicity, to see the variation of the effective volume as a consequence of the entropy increase from quantum fluctuations, we simply consider a subsystem inside the original homogeneous QGP with $V_1<V$, in which energy density $e_1=e_0 + \Delta e$ corresponding to the extra entropy produced in this subsystem, namely, $s_1=s_0+\Delta s$, as in \Fig{fig:demo}. The rest of the system, where entropy density stays the same, has a volume $V_0= V- V_1$. The identification of total energy and entropy leads to an equation,
\be
\label{eq:A1}
s(X_0 e_0 + X_1 e_1) = X_0 s(e_0) + X_1 s(e_1)\,,
\ee
with $X_0=V_0/\Vef$ and $X_1=V_1/\Vef$. Note that $(X_0+X_1)^{-1} = \Vef^{(1)}/V=\Vef^{(1)}/\Vef^{(0)}=1+\Delta \ln \Vef$. With respect to a parameterization of the equation of state, $s(e) \propto e^\kappa$, with $\kappa$ a constant, \Eq{eq:A1} can be solved analytically. By introducing $\varepsilon = \Delta e/e = \Delta \ln e$ and $v = V_1/(V_0+V_1)$ to quantify the inhomogeneity in the system due to the entropy production, one finds the solution,
\be
\label{eq:A2}
\Delta \ln \Vef = \left(\frac{1-v + v(1+\varepsilon)^\kappa}{(1+v \varepsilon)^\kappa}\right)^{-\frac{1}{\kappa - 1}} -1
\ee
Except $\kappa=1$, for which the change of the effective volume becomes undetermined by $v$, for a small increase of entropy or energy, $\varepsilon>0$, as the inhomogeneity vanishes as $v=0$ or $v=1$, the change of the effective volume approaches zero, as expected. For arbitrary finite $v$, \Eq{eq:A2} results in $\Delta \ln \Vef<0$, and reaches maximum with respect to the largest inhomogeneity, i.e., $v=1/2$. 

{\bf Case (II):} We consider a different scenario of entropy production due to initial-state fluctuations, that the quantum fluctuation results in entropy (or energy) produced outside the original system.
We keep the same notions as in Case (I), namely, the volume in which the entropy is produced is $V_1$ such that $v=V_1/V$, and $\Delta e/e=\varepsilon$ corresponds to the extra energy density produced outside the original volume, as shown in \Fig{fig:demo}. Note that now $V_0=\Vef^{(0)}$, and $\Delta \ln \Vef = 1/X_0 -1$, yields,
\be
\Delta \ln \Vef = \left(\frac{1+v \varepsilon^\kappa}{(1+v \varepsilon)^\kappa}\right)^{-\frac{1}{\kappa - 1}} -1
\ee
For absolute homogeneity, namely, $\varepsilon=0$ or $v=0$, $\Delta \ln \Vef=0$. It is easy to show that with respect to a small entropy production, $\Delta \ln \Vef >0 $ for arbitrary $0<v<1$, and the change of effective volume becomes maximum as $v=1$, at which the system becomes most inhomogeneous. 

For the realistic initial state density distributions, as shown in \Fig{fig:entropy}, case (I) applies to systems which are more smooth and connected, so that the extra entropy is likely to be generated inside the original volume, while case (II) characterizes the entropy production in a system that is created initially with a more discrete profile.

\bibliography{references}

\end{document}